\documentclass[twocolumn,showpacs,preprintnumbers,superscriptaddress]{revtex4}
\usepackage{graphicx,bm,times,url}
\usepackage{amsmath}
\graphicspath{{./fig/}{./png/}}
\newcommand{\BoldVec}[1]{\mathchoice%
  {\mbox{\boldmath $\displaystyle     #1$}}%
  {\mbox{\boldmath $\textstyle        #1$}}%
  {\mbox{\boldmath $\scriptstyle      #1$}}%
  {\mbox{\boldmath $\scriptscriptstyle#1$}}%
}
\newcommand{\EQ}{\begin{equation}}
\newcommand{\EN}{\end{equation}}
\newcommand{\EQA}{\begin{eqnarray}}
\newcommand{\ENA}{\end{eqnarray}}

\newcommand{\Eq}[1]{Eq.~(\ref{#1})}

\newcommand{\Fig}[1]{Fig.~\ref{#1}}

\newcommand{\bra}[1]{\langle #1\rangle}

\newcommand{\meanB}{\overline{B}}
\newcommand{\meanJ}{\overline{J}}

\newcommand{\meanBB}{\overline{\bm{B}}}
\newcommand{\meanUU}{\overline{\bm{U}}}

\newcommand{\uu}{\BoldVec{u} {}}

\newcommand{\ww}{\BoldVec{w} {}}

\newcommand{\UU}{\BoldVec{U} {}}

\newcommand{\bb}{\BoldVec{b} {}}

\newcommand{\BB}{\BoldVec{B} {}}

\newcommand{\grav}{\BoldVec{g} {}}

\newcommand{\OO}{\BoldVec{\Omega} {}}

\newcommand{\emf}{\overline{\mbox{\boldmath ${\cal E}$}}{}}{}

\newcommand{\const}{{\rm const}  {}}

\def\urms{u_{\rm rms}}

\newcommand{\yan}[3]{, Astron. Nachr. {\bf #2}, #3 (#1).}

\newcommand{\ymn}[3]{, Mon.\ Not.\ R.\ Astron.\ Soc.\ {\bf #2}, #3 (#1).}

\newcommand{\ynat}[3]{, Nature {\bf #2}, #3 (#1).}

\newcommand{\ypre}[3]{, Phys.\ Rev.\ E {\bf #2}, #3 (#1).}
\newcommand{\yprl}[3]{, Phys.\ Rev.\ Lett.\ {\bf #2}, #3 (#1).}
\newcommand{\yphl}[3]{, Phys.\ Lett.\ {\bf #2}, #3 (#1).}

\newcommand{\yoleb}[3]{, Orig. Life Evol. Biosph. {\bf #2}, #3 (#1).}
\newcommand{\yapj}[3]{, Astrophys. J. {\bf #2}, #3 (#1).}
\newcommand{\yija}[3]{, Int. J. Astrobiol. {\bf #2}, #3 (#1).}
\newcommand{\yab}[3]{, Astrobiol. {\bf #2}, #3 (#1).}

\newcommand{\ygafd}[3]{, Geophys. Astrophys. Fluid Dyn. {\bf #2}, #3 (#1).}

\newcommand{\yjour}[4]{, #2 {\bf #3}, #4 (#1).}

\newcommand{\ybook}[3]{, {\em #2}. #3 (#1).}

\begin{document}
\preprint{NORDITA 2010-98}

\title{Spontaneous chiral symmetry breaking by hydromagnetic buoyancy}
\author{Piyali Chatterjee}
\affiliation{NORDITA, AlbaNova University Center, Roslagstullsbacken 23,
SE-10691 Stockholm, Sweden}

\author{Dhrubaditya Mitra}
\affiliation{NORDITA, AlbaNova University Center, Roslagstullsbacken 23,
SE-10691 Stockholm, Sweden}

\author{Axel Brandenburg}
\affiliation{NORDITA, AlbaNova University Center, Roslagstullsbacken 23,
SE-10691 Stockholm, Sweden}
\affiliation{Department of Astronomy,
Stockholm University, SE 10691 Stockholm, Sweden}

\author{Matthias Rheinhardt}
\affiliation{NORDITA, AlbaNova University Center, Roslagstullsbacken 23,
SE-10691 Stockholm, Sweden}

\date{Received 13 November 2010; \today,~ $ $Revision: 1.162 $ $}

\begin{abstract}
Evidence for the parity-breaking nature of the magnetic buoyancy instability
in a stably stratified gas is reported.
In the absence of rotation, no helicity is produced, but the non-helical state
is found to be unstable to small helical perturbations during the development
of the instability.
The parity-breaking nature of an instability in magnetohydrodynamics
appears to be the first of its kind and has properties similar to those
in chiral symmetry breaking in biochemistry.
Applications to the production of mean fields in galaxy clusters
are discussed.
\end{abstract}


\maketitle

The phenomenon of spontaneous breakdown of chiral symmetry, i.e.,
the bifurcation of an achiral state into two
states with opposite chirality (handedness),
has attracted attention in many different fields of physics, even at the macroscopic level.
A well studied example from hydrodynamics is the
Taylor--Couette flow with counter-rotating cylinders. 
In this case, linear stability analysis reveals the presence of 
two oppositely helical states (spiral vortices) with
identical growth rates.
Experiments and weakly nonlinear analysis show that, depending on the
initial conditions, one of these two
helical states is selected \cite{Hoffetal05}.  
One might suppose that rotation (which is an axial vector) is essential to enable 
the occurrence of such symmetry-degenerate unstable eigenmodes
and hence the breaking of chiral symmetry.
In other words, the conjecture
is that, in the absence of such a vector
helical states will never show up. 
Then the question arises naturally whether another
global
axial vector
can replace rotation in this respect.
Here the magnetic field is a suggestive candidate.

However, more puzzlingly, helical flows are possible even in 
non-rotating, non-magnetic setups as demonstrated in Ref.~\cite{DemSee02}
for one of the asymmetric square patterns of B{\'e}nard convection.
There, the sign of helicity realized in a given cell
is expected to depend
only on the initial condition. 
Yet, when considering that this type of pattern emerges not by a
bifurcation from the trivial
(non-convective)
solution, but from convective rolls
instead, we can again identify an axial vector in the unstable reference
state, namely the vorticity of the rolls.
Note that this vector can, 
at least with respect to a single roll, be considered global.

As helical flows are known to be crucial in producing large-scale
magnetic fields by the so-called $\alpha$ effect \cite{KR80},
the strongest interest in such flows is likely to be found in those
branches of astrophysics, geophysics and planetology which deal with
the origin of galactic, stellar and planetary magnetic fields.  However,
helicity in astrophysical flows is normally not supposed to be caused by
a spontaneous parity breaking.  Indeed, bodies such as the Sun tend to
produce helicity with opposite signs in their two hemispheres.  This is
not surprising, because the Sun is stratified in the radial direction
by its own central gravity $\grav$ and rotates with finite angular
velocity $\OO$.  These two vectors form a pseudoscalar $\grav\cdot\OO$
that changes sign about the equatorial plane and is, by virtue of the
Coriolis force, directly related to the kinetic helicity.

The question now emerges whether finite kinetic and magnetic helicity
can also be produced in the absence of
ingredients that are {\it a priori} given as, for example $\grav$ or $\OO$.
In search of such a process we have to look for an instability
which shows a preference for amplifying helical
velocity (and possibly magnetic) perturbations
in comparison with non-helical perturbations,
but showing of course no preference for one specific sign of helicity.
One particular example is the magnetic buoyancy instability.
It is well known that this instability can produce helical magnetic
and velocity fields owing to the presence of rotation and stratification
\cite{Thelen}.
Nevertheless, it has been explicitly stated \cite{DH11,Thelen}
that, in the absence of rotation or at the equator, the $\alpha$ effect
must vanish because the eigenmodes of the instability are then
degenerate and have opposite helicities.
By contrast, recent work \cite{CMRB10} delivered indications that finite
helicity can emerge even at the equator.
In this paper we elaborate on the possibility that helicity
may be finite, even in the absence of rotation.
We present new simulations
and argue that our findings provide strong evidence for spontaneous
chiral symmetry breaking during the nonlinear phase of the magnetic
buoyancy instability.
In addition to our example, there is now also that of the Tayler
instability which leads to spontaneous chiral symmetry breaking
\cite{GRH11}.
In both cases there is no rotation and helicity emerges in the absence of
quantities that are {\it a priori} given 
such as $\OO$ and either $\grav$ or at least boundary effects.

There is quite a number of cases where spontaneous chiral symmetry
breaking has been discussed.
A somewhat hypothetical example is the production of magnetic helicity
during electroweak baryogenesis \cite{Vac01} and may be
connected with the emergence of baryon asymmetry \cite{JS97}.
Chiral symmetry breaking is also known in biochemistry where it
refers to the consequent selection of one of
two possible forms of biomolecules
(mainly sugars and amino acids)
that are mirror images of each other.
This selection may have taken place during the emergence of life on Earth
\cite{Kondepudi}.
It requires the preferred
replication of molecules of the same handedness,
known as autocatalysis \cite{Frank,Soai}.
Equally important is the effect of mutual antagonism.
This has been identified in the context of DNA
polymerization, where it characterizes the inability to
continue polymerization with
monomers of opposite handedness \cite{Joy84}.
Thus, an initial imbalance in chirality is strongly amplified and
the final selection of one chirality over the other consequently is a result of
random fluctuations in the equipartition of right and left-handed monomers
in a prebiotic mixture  \cite{BLL07}.

The equations governing the agent concentrations in biochemistry
\citep[e.g., Ref.][]{Kondepudi} on the
one hand and the velocity mode amplitudes in hydrodynamics
\citep[e.g.,][]{Fau} on the other show remarkable similarities.
Hence, it could be fruitful to keep the mentioned biochemical processes in
mind when trying to understand bifurcations into helical states in hydrodynamics.
In particular, it deserves interest if something like `mutual antagonism'
can be identified in the nonlinear evolution of two competing eigenmodes
with opposite helicity.

We consider a Cartesian slab
of ideal gas with a stable stratification in the $z$ direction which is
altered by the presence of a narrow horizontal magnetic layer in
magnetostatic equilibrium.
That is, in the layer a fraction of the gas pressure is
substituted by magnetic pressure.
Equilibrium is achieved by reducing the temperature correspondingly, but
keeping the density unchanged, so the magnetic layer is initially not buoyant.
However, the local temperature change in the layer perturbs the
thermodynamic equilibrium.
We solve the compressible hydromagnetic equations in a setup
described in detail in Ref.~\cite{CMRB10},
but focus here on the case without rotation,
$\OO={\bf0}$,
which shows no {\it a priori} preference of positive or negative helicity.

Our computational domain has an aspect ratio $L_x:L_y:L_z=1:3:1$.
The ratio of the thickness of the magnetic layer, $H_B$, to
pressure scale height $H_p$ and $L_z$ is $H_B:H_p:L_z=0.1:0.3:1$.
Our fluid and magnetic Prandtl numbers are equal to 4.
The Lundquist number based on the thickness of the magnetic layer
is $v_{\rm A0}H_B/\eta=1000$,
where $v_{\rm
A0}=B_0/\sqrt{\rho_0\mu_0}$ is the Alfv\'en speed associated with the
initial magnetic field of strength $B_0$ in the $y$ direction,
$\rho_0$ is the initial density
at the bottom of the domain, $\mu_0$ the vacuum permeability, and $\eta$ the magnetic diffusivity.
The initial stratification is a polytrope with index 3,
so density scales to temperature like $\rho\sim T^3$.
We use the fully compressible {\sc Pencil Code} \footnote{
http://www.pencil-code.googlecode.com} for all our calculations
and the number of mesh points is $64^3$.

Considering reflectional symmetries of the basic equations together with
the unperturbed initial state and the boundary conditions, we observe
that reflections at planes $z=\const$ clearly change the system already
due to its stratification.
In contrast, reflections at planes $y=\const$ leave it unchanged while
those at planes $x=\const$ are equivalent with merely changing the sign
of the initial field of the magnetic layer.
Hence there is no essential difference in the behavior of a solution and
a corresponding counterpart obtained by one of the two latter reflections.
In particular, if there are helical eigenmodes of the linearized system
they should occur in pairs with opposite helicities, but equal growth (or
decay) rates.
Further, if there were stable quasi-stationary helical solutions of the
nonlinear system they should again occur in such pairs.

Our results depend decisively on the details of the initial velocity
perturbations, which possess a slight imbalance of random sign
in net kinetic helicity. 
Depending on this sign
we find exponential amplification of positive or negative kinetic
and current helicities during the initial stage of the instability.
An important consequence of this parity breaking is the occurrence of
what is known in dynamo theory as the $\alpha$ effect.
This effect is
crucial for the amplification and sustainment of a
mean magnetic field $\meanBB$. Its
evolution is essentially governed by 
the mean electromotive force, $\emf$, which is the
correlation $\overline{\uu\times\bb}$ of velocity
and magnetic field fluctuations,
$\uu=\UU-\meanUU$ and $\bb=\BB-\meanBB$, respectively,
where the overbars indicate horizontal
($xy$)
averaging.
If $\uu$ lacks reflectional symmetry, and in particular, if it exhibits
handedness, $\overline{\uu\times\bb}$ can have a constituent
proportional to the mean field:
\EQ
(\overline{\uu\times\bb})_i=\alpha_{ij}\circ\meanB_j-\eta_{ij}\circ\meanJ_j,
\label{uxb}
\EN
where the operator ``$\circ$" denotes convolution in space, and in
general also in time, but this will be ignored here.
Note that $\alpha_{ij}$ and $\eta_{ij}$ are tensorial kernels,
the former a pseudo and the latter a true tensor.
The symmetric part of $\alpha_{ij}$ results in the
aforementioned $\alpha$ effect, while that of $\eta_{ij}$
describes what is known as turbulent diffusion.
Given that we are working with horizontal averages,
$\meanBB=\meanBB(z,t)$, we have $\meanB_z=0$ in the absence of
an imposed field.
Therefore, the evolution of $\meanBB$ is governed only by $2\times2$
components of both $\alpha_{ij}$ and $\eta_{ij}$ with 
$\alpha_{xx}$ and $\alpha_{yy}$ being of particular interest.

In Fourier space (with respect to $z$) the convolutions in \Eq{uxb} 
turn into multiplications with tensors that depend on $z$ wavenumber $k$.
However, due to the intrinsic inhomogeneity of our system
they depend at the same time also explicitly on $z$. 
The
functions $\hat{\alpha}_{ij}(z,k)$ and $\hat{\eta}_{ij}(z,k)$
can be directly determined by the so-called test-field method
\cite{Sch05,Sch07}; for specific details see \cite{CMRB10}.
For the purpose of the present paper it suffices to consider
the Fourier constituents with the smallest wavenumber,
$k= \pi/L_z$ and we omit the argument $k$ in what follows.

\begin{figure}
\vspace{-0.4cm}
\begin{center}
\includegraphics[width=\columnwidth]{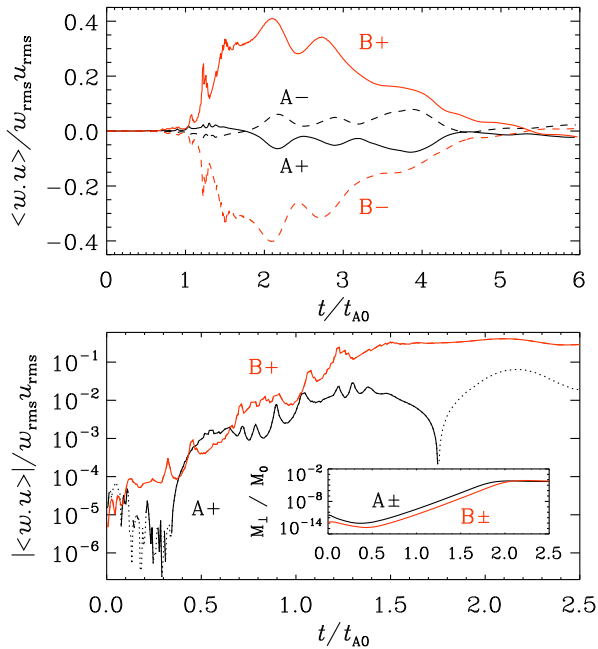}
\caption{\label{fig:kw}
(Color online) Upper panel: Evolution of the normalized kinetic helicity
for runs A$\pm$ and B$\pm$. Lower panel: Same as above but for the
absolute value (dotted) overplotted with a solid line to indicate the
intervals of positive kinetic helicity.
Inset: $M_{\perp}=\bra{B_x^2+B_z^2}$, normalized by the averaged initial
magnetic energy density, $M_0=\bra{B_{y0}^2}$ for the same set of runs.
}\end{center}
\end{figure}

In this paper we discuss two pairs of runs, A$\pm$ and B$\pm$.
For each pair, the initial velocity of one run is an exact mirror image
of that
of the other one with respect to a vertical plane.
Consequently, the two runs have opposite
initial kinetic helicity.
Depending on its sign, the runs are labeled by $+$ or $-$.
In both pairs the initial velocity contains a random pattern of
horizontal eddies with Mach numbers of the order of $10^{-5}$.
In pair A$\pm$,
there are additional random vertical motions with the same Mach number,
whereas in pair B$\pm$ these are set to zero. 
Thus, the pair A$\pm$ differs from pair B$\pm$ in
the magnitude of the initial kinetic helicity, whose normalized value
$\bra{\ww\cdot\uu}/\urms w_{\mathrm{rms}}$
is $\approx 2\times10^{-5}$ for the former and, due to numerical noise
only, $\approx 4\times 10^{-10}$ for the latter.
In the linear phase of the instability the two runs of each pair
yield exactly the same growth rate.
Likewise, the magnetic and kinetic energies in the saturated stage are
nearly the same.
The evolution of the normalized kinetic helicity
is shown in \Fig{fig:kw} for all four runs.
Here, angular brackets denote the average over the full computational domain.
Note that parity breaking becomes obvious after about one Alfv\'en time,
$t_{\rm A0}=L_y/v_{\rm A0}$.
In the lower panel of \Fig{fig:kw} we show the moduli of the kinetic helicity
together with the energy of the
components $B_{x,z}$ generated from the initial $B_y$ by the instability.
Evidently, the evolution of the magnetic energy is nearly the same in all cases.
Note that, along with the kinetic helicity, the current helicity also is
generated with the same growth rate, again with positive
or negative sign, depending on the initial condition.
Furthermore, the final sign of the helicity
does not need to agree with the initial one.
For example, A$+$
had positive helicity initially, but ends up with negative
helicity, undergoing several sign changes during the kinematic stage.
Given that the helicity is a volume integral comprising patches with
opposite signs of $\ww\cdot\uu$,
such sign changes can be understood
in terms of local volume changes of these patches.
The sign change for the A$+$ solution at $t/t_{\rm A0}=1.7$ could then
be explained by a gradual shrinking of a patch with positive helicity
that occupied a much larger fraction at earlier times.
We speculate that
this scenario might be analogous to the case of chiral molecules,
where patches of one handedness would shrink in the direction of the
curvature vector \cite{BM04}.

\begin{figure}

\vspace{-5mm}
\begin{center}
\includegraphics[width=\columnwidth]{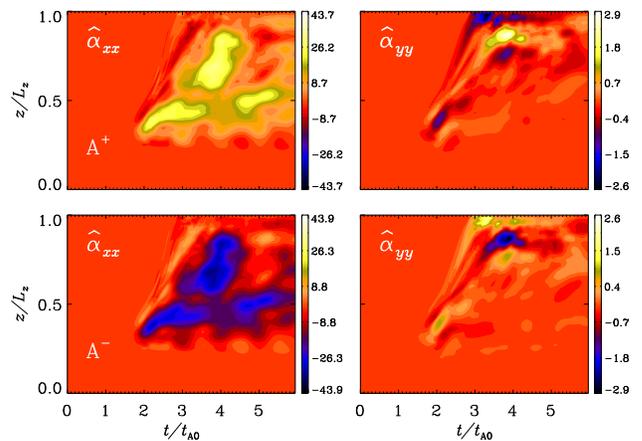}
\caption{\label{fig:alp22}
(Color online) $\hat{\alpha}_{xx}$ and $\hat{\alpha}_{yy}$ as functions
of time and height for runs A$+$ (top) and A$-$ (bottom).
Values scaled by $10^3/v_{\rm A0}$.
}\end{center}
\end{figure}

In \Fig{fig:alp22} we plot the dependence of
$\hat{\alpha}_{xx}$ and $\hat{\alpha}_{yy}$ on $z$ and $t$ for
the runs A$\pm$.
Note that their patterns reflect the
rise of the magnetic layer from its original position at $z=0.3\,L_z$
to the top of the domain at $z=L_z$.
Obviously, within each pair of runs the $\hat{\alpha}_{yy}$ map
of the run with negative initial
helicity, labeled ``$-$'', can be obtained from the one of the run
with positive initial helicity, labeled ``$+$''
to reasonable accuracy by a simple sign inversion.
\begin{figure}
\begin{center}

\vspace{-3mm}
\includegraphics[width=.90\columnwidth]{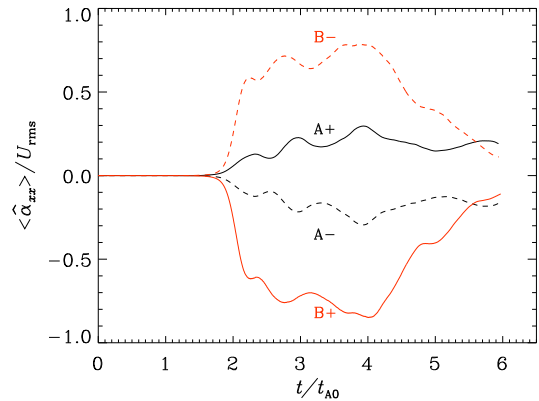}
\caption{\label{fig:alp11}
(Color online) $\bra{\hat{\alpha}_{xx}}$ as a function of time
for runs A$\pm$ and B$\pm$.
}\end{center}
\end{figure}

The component $\hat{\alpha}_{xx}$ has been
found to have the same sign throughout the entire domain
at any instant of time. Hence it is sufficient to consider 
the evolution of the average
$\bra{\hat{\alpha}_{xx}}$ which is shown in \Fig{fig:alp11}
for all four runs.
Its behavior is similar to that of
$\bra{\ww\cdot\uu}$ in \Fig{fig:kw}
which, however, shows a sign 
change at $t/t_{\rm A0}\approx2$, whereas the sign of
$\bra{\hat{\alpha}_{xx}}$ does never change.

The phenomenon of spontaneous parity breaking has been observed 
earlier for hydrodynamic instabilities, in particular those of
the Taylor-Couette flow \cite{Hof}. 
For such instabilities, systems of amplitude equations
of weakly nonlinear, Ginzburg--Landau type 
have already been described in Refs.~\cite{Fau,Hof}.
In all these cases, the helical states, consisting of traveling waves,
emerge due to the presence of rotation which is an axial vector. 
In our case this role is played by the magnetic field.
Deriving nonlinear amplitude equations for the present problem is nontrivial
and will be described elsewhere. 
Here it suffices to say that the amplitude equations in Ref.~\cite{Fau}
show a striking similarity to the equations
that describe homochirality in the biological context \cite{Frank}. 
In particular both sets of equations contain terms of mutual antagonism
which allows one sign of helicity to grow at the expense of the other. 
The analogies between homochirality and the present problem extend even further:
(a) In both cases one can start
      with an initial state with small excess of one sign of
      helicity or chirality,
      but the final state can have the opposite sign \cite{BM04},
      here, e.g., the set of runs A.
(b) In a closed system, which in the biological context implies
that the resources available to the evolving species are
finite, both problems fail to reach a statistically
stationary state \cite{Soai,PB10}.

As eluded to in the beginning, we are not aware of any similar
spontaneous parity breaking instability in magnetohydrodynamics.
A related example is the selection of one type of
non-mirror symmetric crystals during their growth in the presence
of grinding \cite{Noo08,MBT08}.
In the case of the hydromagnetic buoyancy instability presented
here, we can only speculate about possible applications.
In any case we must be thinking of a stably
stratified layer with negligible rotation.
A prime example fitting this description is clusters of galaxies.
Indeed, magnetic and thermal buoyancy effects are invoked to
explain what looks like cold bubbles in such clusters
\cite{Rob04}.

However, the magnetic buoyancy instability might be just one example
showing the parity-breaking
property described here. The magneto-rotational instability (MRI)
could be another one.
In that case there is rotation, but in the absence of stratification
and current density there would still be no
pseudo-scalar constructible and hence no
net helicity, unless the MRI were capable of
spontaneous parity breaking.

Large-scale magnetic fields have been
found in simulations without net helicity including cases with shear,
in particular the MRI \cite{BNST} and the magnetic buoyancy instability
with a sinusoidal shear profile \cite{Cline}.
Magnetic field generation occurs also in the case of a linear shear flow with
isotropic non-helically forced turbulence
Although it is not clear whether in this case an instability (like the
magnetic buoyancy instability) is really required for parity breaking,
it should be noted that there is generally a production of mean vorticity
\cite{Elp07,KMB09}, whose possible importance has been discussed in connection
with dynamos from linear shear flows \cite{Yousef}.
Such a vorticity production would indeed be the result of
a (hydrodynamic) instability, supporting
the idea that an instability might be needed for parity breaking.
An alternative explanation for the origin of large-scale fields in this setup
is the incoherent $\alpha$--shear dynamo \cite{VB97}, where kinetic helicity
is expected to fluctuate about zero.
Revisiting this assertion in the light of the present results
could be a worthwhile effort.

In conclusion, we have found a mechanism that is able to amplify magnetic
and velocity fields with either sign of helicity, but the selection
of one of them occurs during the nonlinear stage where it gives rise
to an $\alpha$ effect under otherwise parity-invariant conditions.
This mechanism might be applicable to clusters of galaxies.
Moreover, it is very likely
that the magnetic buoyancy instability is just a first
example in a class of spontaneous parity-breaking instabilities
in magnetohydrodynamics.

\acknowledgements

The authors thank the
National Supercomputer Centre in Link\"oping and the Center for
Parallel Computers at the Royal Institute of Technology in Sweden.
This work was supported in part by the Swedish Research Council,
grant No.\ 621-2007-4064, and the European Research Council under the
AstroDyn Research Project No.\ 227952.


\end{document}